%% file: webdb.tex
\documentclass{sig-alternate}
\usepackage[hyphens]{url}
\usepackage{framed}
\usepackage{cite}
\usepackage{enumitem}
\usepackage[english]{babel} 
\usepackage{etoolbox}
\makeatletter
\preto{\@verbatim}{\topsep=0pt \partopsep=0pt }
\makeatother

\newcommand{\nop}[1]{}

\toappear{Copyright is held by the author/owner. \\
Seventeenth International Workshop on the Web and Databases (WebDB 2014), 
June 22, 2014 - Snowbird, UT, USA.}

\begin{document}

\title{AutoShard -- Declaratively Managing \break  Hot~Spot Data Objects in NoSQL Document Stores}

%\subtitle{[Extended Abstract]
%\titlenote{A full version of this paper is available as
%\textit{Author's Guide to Preparing ACM SIG Proceedings Using
%\LaTeX$2_\epsilon$\ and BibTeX} at
%\texttt{www.acm.org/eaddress.htm}}}
%
% You need the command \numberofauthors to handle the 'placement
% and alignment' of the authors beneath the title.
%
% For aesthetic reasons, we recommend 'three authors at a time'
% i.e. three 'name/affiliation blocks' be placed beneath the title.
%
% NOTE: You are NOT restricted in how many 'rows' of
% "name/affiliations" may appear. We just ask that you restrict
% the number of 'columns' to three.
%
% Because of the available 'opening page real-estate'
% we ask you to refrain from putting more than six authors
% (two rows with three columns) beneath the article title.
% More than six makes the first-page appear very cluttered indeed.
%
% Use the \alignauthor commands to handle the names
% and affiliations for an 'aesthetic maximum' of six authors.
% Add names, affiliations, addresses for
% the seventh etc. author(s) as the argument for the
% \additionalauthors command.
% These 'additional authors' will be output/set for you
% without further effort on your part as the last section in
% the body of your article BEFORE References or any Appendices.

\numberofauthors{2} %  in this sample file, there are a *total*
% of EIGHT authors. SIX appear on the 'first-page' (for formatting
% reasons) and the remaining two appear in the \additionalauthors section.
%
\author{
% You can go ahead and credit any number of authors here,
% e.g. one 'row of three' or two rows (consisting of one row of three
% and a second row of one, two or three).
%
% The command \alignauthor (no curly braces needed) should
% precede each author name, affiliation/snail-mail address and
% e-mail address. Additionally, tag each line of
% affiliation/address with \affaddr, and tag the
% e-mail address with \email.
%
% 1st. author
\alignauthor
Stefanie Scherzinger\\
       \affaddr{Technical University \break of Applied Sciences Regensburg}\\
       \affaddr{Regensburg, Germany}\\
       \email{\mbox{stefanie.scherzinger@oth-regensburg.de}}
% 2nd. author
\alignauthor
Andreas Thor\\
       \affaddr{Deutsche Telekom}\\
       \affaddr{University of Applied Sciences}\\
       \affaddr{Leipzig, Germany}\\
       \email{thor@hft-leipzig.de}
}

\maketitle

\input{abstract}

% A category with the (minimum) three required fields
%\category{H.4}{Information Systems Applications}{Miscellaneous}
%A category including the fourth, optional field follows...
%\category{D.2.8}{Software Engineering}{Metrics}[complexity measures, performance measures]


%\keywords{ACM proceedings, \LaTeX, text tagging} % NOT required for Proceedings

\input{introduction}

\input{sharding}

\input{using_autoshard}

\input{architecture}

\input{eval}

\input{summary}

{\small
\bibliographystyle{abbrv}
\bibliography{bibliography}
}
\end{document}

%% file: abstract.tex
\begin{abstract}
NoSQL document stores are becoming increasingly popular as backends in web development.
Not only do they scale out to large volumes of data, many systems are even
custom-tailored for this domain: NoSQL document stores like Google Cloud Datastore
have been designed to support massively parallel reads,
and even guarantee strong consistency in updating single data objects.
However, strongly consistent updates cannot be implemented
arbitrarily fast in large-scale distributed systems.
Consequently, data objects that experience high-frequent writes
can turn into severe performance bottlenecks.
In this paper, we present AutoShard, a ready-to-use object mapper for Java applications
running against NoSQL document stores. AutoShard's unique
feature is its capability to 
gracefully shard
hot spot data objects to avoid write contention.
Using AutoShard,  developers can easily handle hot spot data objects
by adding minimally intrusive annotations to their application code.
Our experiments show the significant impact of 
sharding on both the write throughput and the execution time.
\end{abstract}

%% file: introduction.tex
\section{Introduction}

NoSQL document stores are highly appealing in web development,
especially for applications
that  require high scalability and high availability.
Conveniently, NoSQL data stores are readily available with 
established web hosting platforms, such as 
Google App Engine~\cite{sanderson}.
The flexible data model that these systems commonly provide
suits an agile software development style since
the database schema does not have to be designed up front. 
New attributes can be easily added on-the-fly as required by new features.
The simple data access methods (i.e., put() and get(key)) and the 
limited query capabilities are usually sufficient for web applications. 
Moreover, the sheer scalability of these systems is impressive:
Due to a highly distributed architecture,
these systems gracefully handle large amounts of users and data.

NoSQL document stores are capable of scaling out over tens of thousands of nodes.
With such an architecture, arbitrary transactions with ACID semantics  
are usually not feasible, and effects of eventual consistency
come into play.
Yet the application logic often demands strong consistency.
Therefore, systems such as Google Cloud Datastore
that have been designed with web applications in mind,
allow for strongly consistent updates in restricted cases,
e.g.\ when changes affect a single data object only~\cite{google_datastore}.

Yet strongly consistent updates come at the cost 
of slower writes. With Google Cloud Datastore, the supported limit
for writing against a single data object
is merely one update per second~\cite{google_datastore_consistency}.
This can result in {\em write contention}\/, an effect not unique to
this particular system, e.g.~\cite{couchdb_contention}.

Commonly, web applications are read-intensive,
so in the presence of few writes,
this limit may not even be noticed.
However, there are certain features in  web 
applications that are inherently prone to write contention, 
such as a global counter 
recording page visits, or many users hitting a {\em like}\/ button on a 
popular image.

\begin{figure}[t]
\centering
\includegraphics[width=0.8\linewidth]{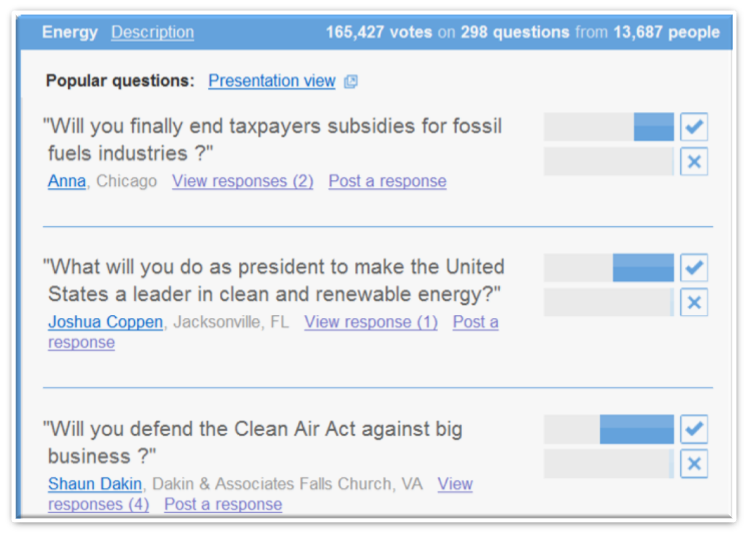}
\caption{Users vote questions up or down in {\em Google Moderator}
during the U.S.\ presidential debates, creating
hot spot data objects.}
\label{fig:moderator}
\end{figure}

\newdef{example}{Example}
\begin{example}
Consider crowd sourcing tools such 
as {\em Google Moderator}
for  ranking user-submitted 
questions (see Figure~\ref{fig:moderator}). 
In 2008, this tool
has been successfully employed in the U.S. presidential debates, with 
one million votes from 20,000 people in just 48~hours%
\footnote{\url{http://en.wikipedia.org/wiki/Google_Moderator}}.
Thanks to a sophisticated implementation, {\em Google Moderator} 
is highly scalable.
However, a naive implementation 
on a backend such as Google Cloud Datastore may not scale:
If the counter tracking positive votes on a question is 
implemented by a {\em single}\/ document,
already tens of users concurrently voting on the same popular question
will cause write contention. 
Ultimately, this triggers runtime errors during peak times.
\end{example}

How to deal 
with so-called {\em hot spot data objects} has already been subject of study 
back in the late~70s and early~80s. Typical use cases in those days were the 
number of available seats on a plane, or the overall balance of bank 
accounts. The idea of exploiting the semantics of data items and 
their transactions has been termed {\em semantics-based transaction  
processing}, and has lead to sophisticated solutions, such 
as the IMS Fast Path system~\cite{REUT82}, or the Escrow method~\cite{escrow}.
A comprehensive survey can be found in~\cite{executive_briefing}.

Yet the challenge of managing hot spot data objects 
in today's web
applications comes with several novel aspects: 

\begin{enumerate}[nosep]
%(1)~%
\item
NoSQL data stores frequently implement {\em optimistic concurrency control}\/, whereas 
solutions designed for relational database systems usually assume pessimistic locking mechanisms to be in place. 

%(2)~%
\item
When NoSQL data stores are used as database-as-a-service, it is
impossible for developers to extend or even customize the functionality of 
their backends. Thus, we need to be able to handle write contention 
on the level of the database application, rather than by
physical database design. 

%(3)~%
\item
Finally, NoSQL data stores commonly do not provide full ACID transaction behavior,
so eventual consistency effects must be considered.
\end{enumerate}

% What is sharding? Why is it difficult?
The established approach in the developer community for managing hot spot
data objects in NoSQL data stores is application-level
{\em sharding} of documents,
e.g.\ as suggested in~\cite{sanderson, datastore_sharding} for Google Cloud Datastore,
\cite{sharding_simpledb} for Amazon SimpleDB, and \cite{couchdb_contention}
for Couchbase.
Like with traditional data fragmentation and allocation
in distributed systems~\cite{pdds},
sharding data objects effectively distributes write requests
across physical nodes.
However, sharding is now managed in the application code: 
For example, instead of 
storing a single counter for voting up a  {\em Google Moderator}\/ question, we maintain~10  shard 
counters. Updates are then performed on a single, randomly chosen shard,
and the sum 
across all shard counters yields the overall vote count.

% Why automatically?
While the idea is intuitive, getting sharding
right is not trivial.
Writing custom sharding code (e.g.\ as exemplified in~\cite{datastore_sharding})
requires a deep understanding of the underlying 
technology and its transactional behavior.
Additionally, this introduces
a new level of complexity in the application code, which in return
 increases the development and testing effort.
This is not to be underestimated when sharding is added in hindsight, to already existing code.
Also, coupling the sharding code with the application logic
enforces the technological lock-in with one particular database provider.

What is missing today is a well-principled 
machinery for sharding that does not amount to major
code refactoring.

%\paragraph{Contributions}%
\smallskip
\noindent
{\bf Contributions.} 
Our main contributions are:

\begin{itemize}[nosep]

\item
We give a systematic overview 
for sharding hot spot data objects in NoSQL document stores.
In particular, we introduce static and dynamic {\em property sharding}\/,
as well as {\em entity group sharding}\/.

\item
We present AutoShard,
a novel Java object mapper specifically designed for NoSQL 
document stores.
AutoShard relieves developers from having to deal with
low-level property sharding, and thus restores a clearer separation
between logical and physical database design.
AutoShard is designed for ease-of-use,
merely requiring simple declarative annotations in Java classes.
Its architecture relies on self-modifying code
to transparently generate sharding code.

\item We evaluate static property sharding with 
AutoShard and thereby demonstrate a significant improvement 
on both the  write throughput 
and the average execution times of writes against hot spot data objects.

\end{itemize}

\smallskip
\noindent
{\bf Organization.}
The remainder of this paper is organized as follows.
Section~\ref{sec:sharding_strategies}
presents various sharding strategies.
We discuss the tradeoffs in sharding,
i.e.\ eliminating severe performance bottlenecks
and thereby trading in eventual consistency effects
that are still tolerable for a large class of practical applications.
In Section~\ref{sec:using_autoshard}, we introduce AutoShard,
our Java object mapper that unburdens developers
from writing custom sharding code. Instead, developers 
conveniently specify which data members are to be sharded
by adding annotations to their Java code.
We present details on the AutoShard architecture
in Section~\ref{sec:architecture}.
We experimentally evaluate our implementation of AutoShard (Section~\ref{sec:eval}),
and conclude with a summary of our work and the directions
for future research.

%% file: sharding.tex
\section{Sharding Strategies}
\label{sec:sharding_strategies}

We next give an overview of common sharding strategies
as established in the developer community.
AutoShard implements two ways of sharding atomic document properties:

\begin{itemize}[nosep]
\item
In {\em static property sharding} (e.g.~\cite{datastore_sharding, couchdb_contention}), 
the number of shards is fixed, whereas
\item
in  {\em dynamic property sharding} (e.g.~\cite{sharding_simpledb}),
the number of shards grows on demand.
\end{itemize}
Property sharding is applicable to a large class of NoSQL 
document stores.
We illustrate the idea behind these approaches, and further discuss
{\em entity group sharding}\/, where sharding is applied to groups of documents.
Entity groups are a feature specific to Google Cloud Datastore~\cite{sanderson}
and its underlying software layer, the Google-internal Megastore~\cite{megastore}.

\subsection{Static Property Sharding}
\label{sec:static_property_sharding}

Let us resume our discussion of building a scalable voting application.
An example of a voting question is shown below in JSON format.
We refer to persisted objects as {\em entities}\/.

\begin{center}
{\small
\begin{verbatim}
{"kind" : "Question", "id" : 42,
 "question" : 
      "How do you plan to improve public education?",
 "author" : "Phil R",
 "responses" : [ 
    {"response" :
     "i have earned $1048 dollars just by ad clicks",
     "author" : "twodollarclick"} ],
 "votes" : 76} 
\end{verbatim}
}
\end{center}

Each entity
 is assigned a {\em kind}\/, which is simply a classification
of the entity as a question in this case. Each entity has a unique identifier
and further {\em properties}\/ (c.f.\ attributes). Properties may 
be atomic, multi-valued, structured, and even nested (e.g.\ like the list of responses above).

The rate at which users may vote on this question
is physically limited. In systems such as
Google Cloud Datastore, 
only a minimum write throughput of one write per second per entity is guaranteed
(with 5 to 10 concurrent writes achievable on average~\cite{gaewrites}).
Yet a controversial question is likely to receive concurrent votes.
Write contention then
causes runtime errors, and ultimately, results in data loss,
since not all updates can be persisted.

The recommended approach is to {\em  shard}\/ property {\tt votes},
creating~$n+1$ entities instead.
One entity stores the question without the {\tt votes}-property,
we refer to it as the {\em main entity}.
The value of the {\tt votes}-property is distributed over~$n$ single entities,
the {\em shards}\/.
This is shown below.
The first shard with identifier \mbox{\tt 42-1} stores the original value of the {\tt votes}-property,
whereas the {\tt shard\_votes}-property in all other shards has been set to zero.
We can  always obtain the total number of votes for a given question
by computing the sum over the \mbox{\tt shard\_votes} across all~$n$ shards.

\begin{center}
{\small
\begin{verbatim}
  {"kind" : "Question", "id" : 42,
   "question" : 
      "How do you plan to improve public education?",
   "author" : "Phil R",
   "responses" : [ 
      {"response" :
       "i have earned $1048 dollars just by ad clicks",
       "author" : "twodollarclick"} ]}

  {"kind" : "Shard",  "id" : "42-1", 
   "question" : "42", "shard_votes" : 76}

  {"kind" : "Shard",  "id" : "42-2", 
   "question" : "42", "shard_votes" : 0}      ...

  {"kind" : "Shard",  "id" : "42-n", 
   "question" : "42", "shard_votes" : 0}
\end{verbatim}
}
\end{center}

Whenever the question is voted on, a single shard is picked at random,
 \mbox{\tt shard\_votes} is incremented, and the shard is persisted
again. This can usually be executed as an atomic action,
and effectively distributes concurrent writes across the~$n$ shards,
rather than all concurrent updates affecting a single entity. 
Since addition of integers is commutative and
associative, this is mathematically sound.

{\bf Tradeoffs.}
By sharding the hot spot counter, we have eliminated a crucial scalability
bottleneck in our voting application. As we show in Section~\ref{sec:eval},
sharding significantly improves the write throughput on single entities,
while keeping the average transaction time within acceptable bounds.
Yet sharding hast two inherent drawbacks, owing to the particularities
common to many NoSQL document stores:
\begin{enumerate}[nosep]

\item 
Range queries over shards may not be supported.

\item
The computation of the total number of votes
may show eventual consistency effects.

\end{enumerate}

Let us elaborate on drawback~(1):
NoSQL document stores commonly provide very restricted
query languages. For instance, Google Cloud Datastore
would support the following query 
over the original, unsharded question entity:

\begin{center}
{\small
\begin{verbatim}
      select * from Question where votes > 50
\end{verbatim}
}
\end{center}

Yet the query language is not expressive enough
to compute the equivalent query in the presence of shards.
Consequently, developers need to write custom code
to retrieve these questions.
On the good side, queries filtering over unsharded properties
can still be expressed.

(2)~We next consider compromises in consistency.
In many NoSQL data stores, updating a single entity is
a strongly consistent action (c.f.~\cite{nosql_distilled}).
Let us assume that shards~\mbox{\tt 42-2} and~\mbox{\tt 42-3} from our example have been updated concurrently:

\begin{center}
{\small
\begin{verbatim}
  {"kind" : "Shard",  "id" : "42-2", 
   "question" : "42", "shard_votes" : 1}

  {"kind" : "Shard",  "id" : "42-3", 
   "question" : "42", "shard_votes" : 1}
\end{verbatim}
}
\end{center}

At this point, the total number of votes reaches~$78$. 
Let us try to retrieve this value. Since the query language
is not expressive enough to aggregate across several entities,
we first issue a query to fetch all shards for question~\mbox{\tt 42}:

\begin{center} % used only so that the text is not too squeezed
{\small
\begin{verbatim}
      select * from Shard where question = 42
\end{verbatim}
}
\end{center}

This query is evaluated across a large cluster of nodes,
and thus may not return a strongly consistent result.
Next, we programmatically aggregate over the \mbox{\tt shard\_votes}.
Due to the effects of eventual consistency, repeatedly executing the query at time of the updates
may return the stale results~$76$ or~$77$, and eventually will return
the consistent value~$78$.

In an application such as the voting app, temporarily stale results are tolerable,
as long as queries return the consistent state by the time that the result is 
to be utilized (e.g.\ when the presidential debate actually begins).
Therefore, sharding trades strong consistency 
for scalability in terms of concurrent writes.
This is a valid tradeoff for applications 
where we are mainly interested in a ballpark number (e.g.\ counting
the number of visitors to a website),
and where the order of updates does not matter (unlike an auctioning site, for example).
After all, the alternative is an application that suffers from runtime errors and data loss 
at peak times.

\subsection{Dynamic Property Sharding}

We now introduce an alternative approach,
which we refer to as dynamic property sharding. The previous discussion of tradeoffs
applies here as well.
In our running example, we 
start with a {\em single}\/ shard where the  property
\mbox{\tt shard\_votes} is set to the original value of {\tt votes}.

\begin{center}
{\small
\begin{verbatim}
  {"kind" : "Shard",  "id" : "473", 
   "question" : "42", "shard_votes" : 76}
\end{verbatim}
}
\end{center}

For each user who increments the counter,
we add a new shard and let the NoSQL data store assign a unique key.
Thus, when two users increment the counter concurrently, 
two new shards with \mbox{\tt shard\_votes}=$1$ are added:

\begin{center}
{\small
\begin{verbatim}
  {"kind" : "Shard",  "id" : "119", 
   "question" : "42", "shard_votes" : 1}

  {"kind" : "Shard",  "id" : "236", 
   "question" : "42", "shard_votes" : 1}
\end{verbatim}
}
\end{center}

Even under immense write load,
increments can be executed without {\em any}\/ concurrent writes against
a single entity.
This comes at the cost of higher storage requirements.
Again, the total shard value is computed by aggregating over 
all shards. This may temporarily yield stale results, again due to eventual consistency effects.
An independent batch process, e.g., run nightly or when the
system is under less load, compacts the shards that have accumulated.
This reduces the number of shards,
as well as the storage costs.
Since there is a single thread writing (or rather, deleting) the shards,
this does not cause write contention.
Dynamic property sharding scales more gracefully under peak loads,
yet amounts to a considerable implementation effort,
involving background batch processes or MapReduce jobs.

\subsection{Entity Group Sharding}

Entity groups are a particular feature of Google Cloud Datastore~\cite{google_datastore} 
and Megastore~\cite{megastore}.
Entities can be arranged in groups by defining
a hierarchy between entities.
By physically co-locating the entities inside a group, the system can guarantee
ACID updates within the scope of the group. 

Different from our original data design of a question with nested responses,
we can store the responses in the same group as the question.
Below, property {\tt parent-id} references the question as
the root of the hierarchy.

\begin{center}
{\small
\begin{verbatim}
  {"kind" : "Question", "id" : 42,
   "question" : 
      "How do you plan to improve public education?",
   "author" : "Phil R",
   "votes" : 76}

  {"kind" : "Response", "id" : 47, "parent-id" : 42,    
   "response" :
       "i have earned $1048 dollars just by ad clicks",
   "author" : "twodollarclick"}

  {"kind" : "Response", "id" : 67, "parent-id" : 42,    
   "response" : "Crucial for our future",
   "author" : "Stan S"}
\end{verbatim}
}
\end{center}

As with single entities, Datastore limits the number of concurrent writes
against an entity group.
Several users responding to a question in a heated debate
thus turn the group into a hot spot data object.
In entity group sharding, we consequently shard entity groups,
now distributing writes over several groups.
To restore all entities from the original group,
we compute the union of entities from across several groups.
This improves the rate of successful writes, at the cost of 
making certain atomic updates impossible.%
\footnote{Google Cloud Datastore only allows ACID transactions
involving up to five entity groups~\cite{sanderson}.}

%% file: using_autoshard.tex
\section{The AutoShard Object Mapper}
\label{sec:using_autoshard}

\begin{figure}[t]
\centering
\begin{framed}
\small
\begin{verbatim}
@Entity class Question {
  @Id private int id;
  private String question;
  private String author;
  private List<Response> responses;
	
  @Shardable (neutral=0, shards=10)
  private int votes = 0;

  @ShardMethod
  public void voteUp() {
    this.votes++;
  }

  @ShardFold
  public static int foldVotes(int x, int y) {
    return x + y;
  }
  /* ... not showing getters and setters ... */
}
\end{verbatim}

\end{framed}

\vspace{-5mm}
\caption{Java class with AutoShard annotations.}
\label{fig:class_counter}
\end{figure}

The main focus of our work is 
a novel object mapper framework for automatically rewriting annotated Java classes
with property sharding. 
Like other object mappers,
(e.g.~\cite{datastore_jdo,datastore_jpa, objectify, morphia}), 
AutoShard takes care of the mundane marshalling of 
persisted entities into Java objects and back,
thus greatly simplifying application development.
Just like established 
object mappers, AutoShard relies on Java language metadata annotations.

We show how AutoShard helps with our running example, the voting app. 
The Java class from Figure~\ref{fig:class_counter}
represents a  {\tt Question}
that can be voted up.
In the following, we resolve write contention
by property sharding.

As is customary with object mappers, 
the annotation {\tt @Entity} specifies that an instance of class {\tt Question}
is to be persisted as an entity. The annotation {\tt @Id} marks
the unique key of the persisted entity.

The mapping of an instance of class {\tt Question} 
onto a persisted entity is straightforward. 
As discussed previously,
a {\em single}\/ entity 
is a performance bottleneck with multiple users  voting concurrently
on the same question.
To solve this problem by sharding, we merely 
add annotations. The annotation {\tt @Shardable} 
specifies that the class member {\tt votes} is to be sharded.%
\footnote{Figure~\ref{fig:class_counter} shows the simple case of a single
sharded class member. Naturally, the syntax of AutoShard annotations also allows for several 
data members to be sharded.}
When processing shards, the method annotated with {\tt @ShardMethod} 
will be applied to a single shard, rather than the
global value of the votes counter.
We could even declare several sharding functions (e.g., to increment and 
decrement votes). 
Since in this example the shard method is incrementation, we specify zero as the neutral element
(see \mbox{\tt neutral=0}). 
This information is exploited in initializing new shards.
Further, we request static property sharding with ten 
shards (specifying {\tt shards=10}).
If no shard limit is specified, AutoShard shards dynamically.

With annotation {\tt @ShardFold}, we declare the static
function {\tt foldVotes} as the folding function.
This function is called for aggregating over all shards.
We may specify even more complex folding operations,
as long as they are commutative and associative.
It is the responsibility of the developers to correctly annotate their 
Java classes.

%% file: architecture.tex
\section{The AutoShard Architecture}
\label{sec:architecture}

The AutoShard object mapper is, to our knowledge,
the first Java object mapper to shard properties based on
simple annotations. Our approach relies on self-modifying code
by blending Java code with Groovy technology~\cite{groovy}.
Groovy is a dynamic language that runs in the JVM
and smoothly inter-operates with Java code.
Further, Groovy allows us to annotate code structures for transformations
in the abstract syntax tree (AST) during compilation.

\begin{figure}
\centering
\includegraphics[scale=0.27]{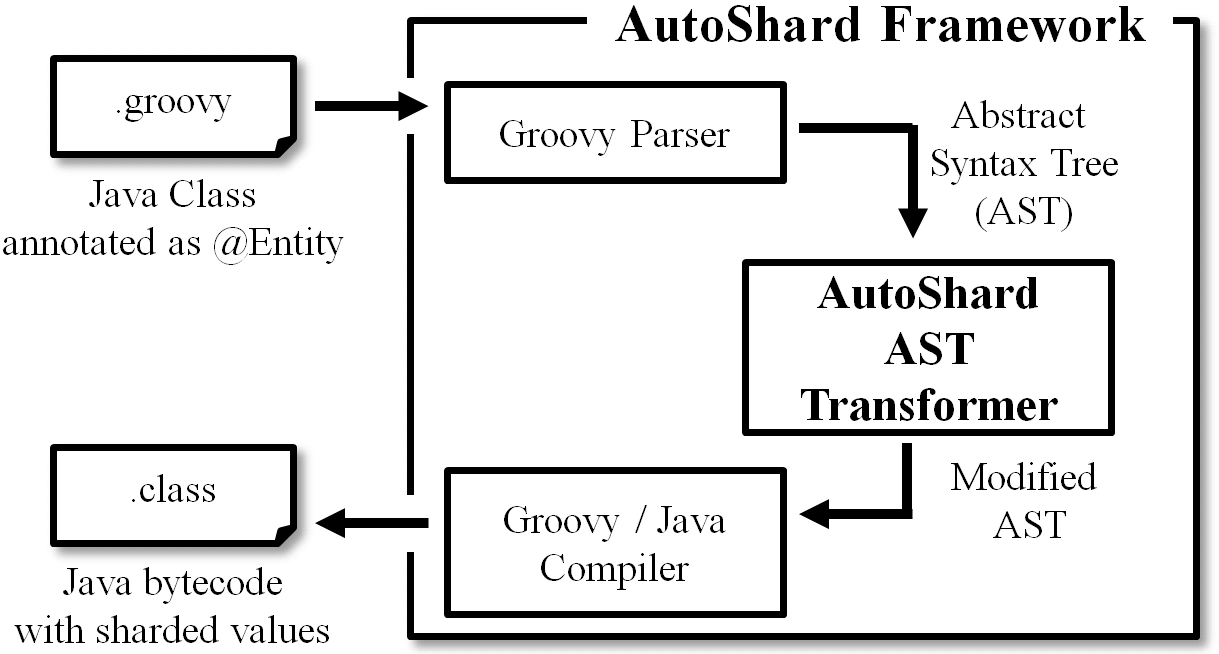}
\caption{The AutoShard framework architecture for compiling
annotated Java classes.}
\label{fig:architecture}
\end{figure}

Figure~\ref{fig:architecture} shows the architecture of the AutoShard 
framework. A Java class with AutoShard annotations 
serves as input. The 
Groovy parser produces an AST and our AutoShard AST transformer restructures 
this tree. Class members annotated as {\tt @Shardable}, as well as 
the sharding and the folding method, are now transformed.

We consider the modifications required for the class from
Figure~\ref{fig:class_counter}.
When compiling for Google Cloud Datastore,
AutoShard's Groovy-based compiler generates the Java class shown in Figure~\ref{fig:class_counter_modified}.
For the sharded property {\tt votes}, compilation introduces a new (private) attribute
 {\tt shard\_votes} that stores a single shard value.
The body of user-defined method {\tt voteUp} is transferred to a private method,
and the original method is replaced
as shown in Figure~\ref{fig:class_counter_modified}.
This new implementation calls the original function both for the shard value ({\tt shard\_votes}) and for the actual value ({\tt votes}). 
Since the signatures of the class methods do not change, 
the remaining application code 
need not be adapted.

\begin{figure}[t]
\centering
\small
\begin{framed}
\begin{verbatim}
class Question {
  private int votes;     // the aggregated value
  private int shard_votes; // single shard value

  // internal method with body of 
  // original voteUp method 
  private void _voteUp() {
    this.votes++;
  }

  // new voteUp method 
  public void voteUp() {
    int tmp = votes;
    votes = shard_votes;
    _voteUp(); // updating the shard value
    shard_votes = votes;

    votes = tmp;
    _voteUp(); // updating the aggregated value
  }

 // the @ShardFold method
  public static int foldVotes(int x, int y) {
    return x + y;  
  }
  /* ... not showing unsharded class members, 
     getters, and setters ... */
}
\end{verbatim}
\end{framed}
\vspace{-5mm}
\caption{The modified Java class \texttt{Question} as generated by AutoShard during compilation.}
\label{fig:class_counter_modified}
\end{figure}

At runtime, we use the modified {\tt Question} class during loading, updating, and saving entities:

{\bf Loading:} 
When a new instance of a {\tt Question} is loaded, AutoShard 
retrieves the main entity to map all unsharded 
class members. For the sharded class member {\tt votes} it reads 
all shards and generates {\em two} data members. 
First, the (regular) data member {\tt votes} is initialized to the 
aggregated shard value. AutoShard uses the {\tt 
@ShardFold} method ({\tt foldVotes} for class {\tt Question}) to aggregate over all shards. 
Second, the (internal) data member {\tt shard\_votes} is 
initialized to the neutral element zero.

{\bf Updates:} 
When shard method {\tt voteUp} is invoked for updating
the counter value, the update is executed on 
both the (regular) data member {\tt votes}
as well as the (internal) data member {\tt shard\_votes}.
This ensures that whenever the application code accesses 
{\tt votes}, it sees the expected value.

{\bf Saving:}
When entity {\tt Question} is persisted
after changes have been made,
AutoShard first updates the main entity. 
For the sharded class member 
a random shard is loaded from storage, as shown in Figure~\ref{fig:class_autoshard_save}. 
Its value is updated by invoking the {\tt @ShardFold}  
({\tt foldVotes} for class {\tt Question})
method on the loaded shard value and on {\tt shard\_votes}.
The shard is persisted
within a nested transaction, so that we do not 
interfere with any transactions that 
may be running in the remaining code.
Since the sharded value is re-set to the neutral element,
it will capture future updates.
Note that the regular property {\tt votes} still holds
the current value. 

% old version
\nop{
\begin{verbatim}
public void save(Question q) {  
  DataStore.put(q);	// save unsharded properties
  BEGIN TRANSACTION
    // read ONE shard value at random    
    shard = DataStore.getShard(q);	
    // fold shard value with object shard property
    // using @ShardMethod
    fold = q.foldVotes(shard.value, q.shard_votes); 
    // save updated shard value
    shard.value = fold;
    DataStore.put(shard);
    // re-initialize object shard property
    q.shard_votes = 0; // 0 = neutral element
  END TRANSACTION
}
\end{verbatim}
}

\begin{figure}[t]
\centering
\small
\begin{framed}
\begin{verbatim}
public void save(Question q) {  
 DataStore.put(q);     // save the main entity
 BEGIN TRANSACTION
  // read ONE shard value at random    
  shard = DataStore.getRandomShard(q);

  // fold shard value with object shard property
  // using the @ShardFold method
  shard.shard_votes = 
  Question.foldVotes(shard.shard_votes, q.shard_votes); 

  // save the updated shard
  DataStore.put(shard);

  // re-initialize local shard_votes data member
  q.shard_votes = 0; // 0 = neutral element
 END TRANSACTION
}
\end{verbatim}
\end{framed}
\vspace{-5mm}
\caption{Pseudo-code for saving sharded \texttt{Question} entities with AutoShard.}
\label{fig:class_autoshard_save}
\end{figure}

Persisting entities
and retrieving them by key are the main 
building blocks for web applications when interacting with the NoSQL backend.
NoSQL document stores also provide basic query languages. For example, 
Google Cloud Datastore allows queries on entities of the same kind
(or type)
using simple property filters. Queries on 
unsharded properties or entities are not affected by AutoShard compilation. Hence,
they can be run without changes (c.f.\ Section~\ref{sec:static_property_sharding}).

%% file: eval.tex
\section{Evaluation}
\label{sec:eval}

We investigate the runtime benefits of sharding with AutoShard.
We have implemented AutoShard with property sharding for
Google Cloud Datastore,
a commercial NoSQL document store 
handling  6.3 trillion daily requests%
\footnote{Quoting Urs H\"olzle in his keynote at {\em Google Cloud Platform Live 2014},
available online at \url{https://cloud.google.com/events/google-cloud-platform-live/}.}.
Our evaluation scenario deals with a Java implementation of a 
voting tool in the style of Google Moderator. The 
application is hosted on Google App Engine.

\begin{figure}[t]
\centering
\includegraphics[scale=0.49]{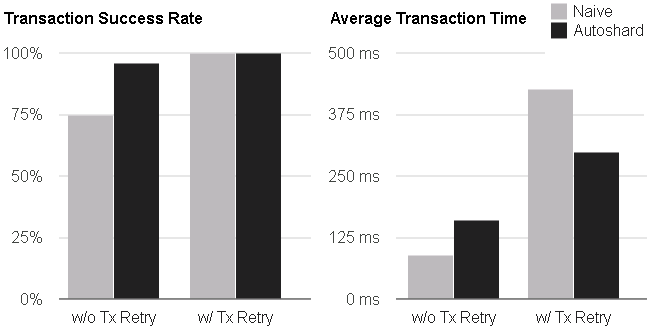}
\caption{Evaluation of static property sharding  where 2,000 users vote on 16 questions. 
A naive implementation shows an unacceptable failure rate of over 25\%.
AutoShard reduces the failure rate down to 4\% (using 16 shards). 
When adding a transaction retry mechanism to ensure a 100\% success rate,
AutoShard significantly reduces the average transaction time compared to the naive implementation.}
\label{fig:eval}
\end{figure}

We start with a naive implementation that does not take  
precautions for handling concurrent writes.
A shell script simulates an increasing number of 
users voting on popular questions, e.g., 75 voting requests per second are equally spread 
across 16 questions. This causes 
write contention on the level of persisted entities.
As seen in Figure~\ref{fig:eval}, for this naive implementation without any transaction
retries (naive ``w/o Tx retry''),  25\% of the transactions fail due to write contention.
This failure rate is obviously unacceptable for real world web applications.

We then repeat the experiment with a sharded version, where we have added 
the AutoShard annotations
from Figure~\ref{fig:class_counter}
to the code and have recompiled the application.
This time, the application 
experiences only 4\% of failed requests (see AutoShard ``w/o TxRetry'').
This improvement is due to the fact that 
write contention is reduced by distributing writes across multiple shards.
However, 
a failure rate of 4\% can still be considered alarming.
Note that in this experiment, we can observe a slight increase in the average transaction time,
due to the overhead imposed by sharding.

To ensure that all votes are indeed persisted,
we add transaction retries (``w/TxRetry''),
so transactions retry until they succeed.
This obviously increases 
execution time.
We repeat the experiment with the naive implementation,
as well as with the code generated by AutoShard, and
visualize the performance results in Figure~\ref{fig:eval}. 
Both versions show a 100\% success rate,
yet the average transaction time for the sharded version is clearly superior
(300ms vs.\ 430ms).

This brief evaluation scenario confirms the usefulness of AutoShard, i.e., Autoshard is
capable of gracefully sharding hot spot data objects. 
An extended evaluation will compare static and dynamic sharding and will analyze 
the impact of the number of shards on the transaction time.

%% file: summary.tex
\section{Summary and Future Work}

In this paper we have presented AutoShard, a ready-to-use object mapper for Java applications
running against NoSQL document stores. In addition to mapping Java objects
to persisted entities, AutoShard is capable of sharding properties
so that hot spot data objects can be managed gracefully.
This form of application-managed sharding ties in with the long tradition
of efforts to avoid write contention over hot spot data objects (c.f.~\cite{executive_briefing}).

A main strength of AutoShard is the ease
with which data objects may be sharded, namely
by merely annotating the Java code.
We have demonstrated the merits of AutoShard
by contrasting the performance of a naive implementation
of a realistic web application with a sharded version generated by AutoShard.
Our experiments show the significant impact of 
property sharding on the throughput of write requests.

In our future work, we are investigating a generic approach
to property and entity group sharding that is not specific to a certain data store.
Programmers should be able to describe all relevant data store properties (e.g., consistency model, ACID guarantees)
so that AutoShard can implement suitable sharding strategies. 
We will examine how to automatically identify 
properties that require sharding as well as to automatically 
determine a suitable number of shards.
We are currently extending AutoShard so that queries involving sharded properties or entities are handled transparently
(whenever possible).
AutoShard is scheduled to be made available as open source software.